\documentclass[aps,twocolumn,prd,a4paper,superscriptaddress,floatfix]{revtex4-1}
\usepackage{graphicx}
\begin{document}

\newcommand{\be}{\begin{equation}}
\newcommand{\ee}{\end{equation}}
\newcommand{\bq}{\begin{eqnarray}}
\newcommand{\eq}{\end{eqnarray}}
\newcommand{\bsq}{\begin{subequations}}
\newcommand{\esq}{\end{subequations}}
\newcommand{\bc}{\begin{center}}
\newcommand{\ec}{\end{center}}
\newcommand\lapp{\mathrel{\rlap{\lower4pt\hbox{\hskip1pt$\sim$}} \raise1pt\hbox{$<$}}}
\newcommand\gapp{\mathrel{\rlap{\lower4pt\hbox{\hskip1pt$\sim$}} \raise1pt\hbox{$>$}}}

\title{Effects of Biases in Domain Wall Network Evolution}
\author{J. R. C. C. C. Correia}
\email{Jose.Correia@astro.up.pt}
\affiliation{Centro de Astrof\'{\i}sica, Universidade do Porto, Rua das Estrelas, 4150-762 Porto, Portugal}
\affiliation{Faculdade de Ci\^encias, Universidade do Porto, Rua do Campo Alegre 687, 4169-007 Porto, Portugal}
\author{I. S. C. R. Leite}
\email{Ines.Leite@astro.up.pt}
\affiliation{Centro de Astrof\'{\i}sica, Universidade do Porto, Rua das Estrelas, 4150-762 Porto, Portugal}
\affiliation{Faculdade de Ci\^encias, Universidade do Porto, Rua do Campo Alegre 687, 4169-007 Porto, Portugal}
\author{C. J. A. P. Martins}
\email{Carlos.Martins@astro.up.pt}
\affiliation{Centro de Astrof\'{\i}sica, Universidade do Porto, Rua das Estrelas, 4150-762 Porto, Portugal}

\date{6 June 2014}
\begin{abstract}
We study the evolution of various types of biased domain wall networks in the early universe. We carry out larger numerical simulations than currently available in the literature and provide a more detailed study of the decay of these networks, in particular by explicitly measuring velocities in the simulations. We also use the larger dynamic range of our simulations to test previously suggested decay laws for these networks, including an ad-hoc phenomenological fit to earlier simulations and a decay law obtained by Hindmarsh through analytic arguments. We find the latter to be in good agreement with simulations in the case of a biased potential, but not in the case of biased initial conditions. 
\end{abstract}
\maketitle

\section{\label{intr}Introduction}

Topological defects necessarily form at phase transitions in the early universe \cite{KIBBLE,VSH}. The majority of the work on defects tends to focus on cosmic strings---and, more recently, superstrings---which are usually benign in terms of their cosmological consequences \cite{KIBCOP,NEWTON}. Domain walls have been comparatively neglected since they are subject to tighter constraints, most notably the so called Zel'dovich bound \cite{ZEL}, and therefore their possible roles in some cosmological mechanisms are significantly reduced \cite{IDEAL,DEVAL}

Nevertheless, domain walls can provide an interesting source of toy models, where it may be easier to understand the various dynamical mechanisms at play in defect network evolution, and this knowledge can subsequently be used for tackling more complicated (but also more realistic) scenarios. This is particularly true in the case of field theory numerical simulations, where domain walls can be described by a single scalar field (although models with further degrees of freedom can certainly be studied)

In the simplest cosmological scenarios, it is well known \cite{VSH} that the attractor solution for the evolution of domain wall networks is a linear scaling solution, where the walls have constant RMS velocities and the characteristic lengthscale of the network (which one may think of as a correlation length, a typical separation between the walls, or a typical curvature radius) grows linearly, \emph{i.e.}, as fast as allowed by causality. In particular, this has been recently confirmed in very high resolution simulations \cite{Leite1,Leite2}.

In the current work we quantify whether (and, if so, how) the linear scaling solution breaks down in several alternative scenarios, where the standard initial conditions are biased in one of several ways. Specifically we consider the cases of anisotropic walls \cite{Fossils}, biased initial conditions \cite{Coulson,Larsson}, and a biased potential \cite{Gelmini,DEVAL}. In some of these we confirm earlier work, though our results are based on simulations with significantly higher resolution and dynamical range, and we also present, for the first time, measurements of the evolution of the averaged wall network velocities in these scenarios. Throughout this paper we shall use natural units with $c=\hbar=1$.

\section{\label{dwevol} Standard domain wall evolution}

In this work we will mostly be interested in flat homogeneous and isotropic Friedmann-Robertson-Walker (FRW) universes. A scalar field $\phi$ with Lagrangian density
\begin{equation}
\mathcal{L}=\frac{1}{2}(\partial_\mu\phi)(\partial^\mu\phi)-V(\phi)\,,
\label{action1}
\end{equation}
where we will take $V(\phi)$ to be a $\phi^{4}$ potential with two degenerate minima, such as
\begin{equation}
V(\phi)=V_0\left({\frac{\phi^{2}}{\phi_{0}^{2}}}-1\right)^{2}\,,
\label{potential}
\end{equation}
will have domain wall solutions, with the height of the potential barrier and surface tension being respectively
\begin{equation}
V_0=\frac{\lambda}{4}\phi_{0}^4\,
\end{equation}
\begin{equation}
\sigma\sim\sqrt{\lambda}\phi_{0}^3\,,
\end{equation}
while the wall thickness is
\begin{equation}
\delta\sim\frac{\phi_{0}}{\sqrt{V_0}}\sim(\sqrt{\lambda}\phi_{0})^{-1}\,.
\end{equation}

By the standard variational methods we obtain the field equation of motion (written in terms of physical time $t$)
\begin{equation}
{\frac{{\partial^{2}\phi}}{\partial t^{2}}}+3H{\frac{{\partial\phi}}{\partial t}}-\nabla^{2}\phi=-{\frac{{\partial V}}{\partial\phi}}\,.\label{dynamics}
\end{equation}
where $\nabla$ is the Laplacian in physical coordinates, $H=a^{-1}(da/dt)$ is the Hubble parameter and $a$ is the scale factor, which we assume to vary as $a \propto t^\lambda$; in particular, in the radiation era $\lambda=1/2$, while in the matter era $\lambda=2/3$.

In order to numerically simulate this model we apply the procedure of Press, Ryden and Spergel \cite{Press}, modifying the equations of motion in such a way that the thickness of the domain walls is fixed in co-moving coordinates. One expects that this will have a small impact on the large scale dynamics of the domain walls, since a wall's integrated surface density (and surface tension) are independent of its thickness. In particular, this assumption should not affect the presence or absence of a scaling solution \cite{Press}, provided one uses a minimum thickness \cite{MY1}---we will briefly revisit this issue below. (For a detailed discussion of analogous issues in the context of cosmic strings see \cite{Moore}.)

In the PRS method, equation (\ref{dynamics}) becomes:
\begin{equation}
{\frac{{\partial^{2}\phi}}{\partial\eta^{2}}}+\alpha\left(\frac{d\ln
a}{d\ln\eta}\right){\frac{{\partial\phi}}{\partial\eta}}-{\nabla}^{2}\phi=
-a^{\beta}{\frac{{\partial
V}}{\partial\phi}}\,.\label{dynamics2}
\end{equation}
where $\eta$ is the conformal time and $\alpha$ and $\beta$ are constants: $\beta=0$ is used in order to have constant co-moving thickness and $\alpha=3$ is chosen in 3D to require that the momentum conservation law of the wall evolution in an expanding universe is maintained \cite{Press}. In fact we have simulated networks with various values of the damping coefficient $\alpha$.

Equation (\ref{dynamics2}) is then integrated using a standard finite-difference scheme. The usual choice of initial conditions assumes $\phi$ to be a random variable between $-\phi_{0}$ and $+\phi_{0}$ and the initial value of $\partial\phi/\partial\eta$ to be zero. This will lead to large energy gradients in the early time steps of the simulation, and therefore the network will need some time (which is proportional to the wall thickness) to wash away these initial conditions. The conformal time evolution of the co-moving correlation length of the network $\xi_c$ (specifically $A/V\propto \xi_{c}^{-1}$, $A$ being the comoving area of the walls) and the wall velocities (specifically $\gamma v$, where $\gamma$ is the Lorentz factor) are directly measured from the simulations, using techniques previously described in \cite{MY2}.

Our main concern here is with a diagnostic for scaling. One looks for the best fit to the power laws
\begin{equation}
\frac{A}{V}\propto\rho_{w}\propto\frac{1}{\xi_{c}}\propto\eta^{\mu}\,,
\label{fit1}
\end{equation}
\begin{equation}
\gamma v\propto\eta^{\nu}\,;
\label{fit2}
\end{equation}
for a scale-invariant behavior, we should have $\mu=-1$ and $\nu=0$. The behavior of the scaling exponent for the network's kinetic energy (or velocity), $\nu$, has not been previously explored in the non-standard scenarios we will be considering.

\section{\label{fossils}Anisotropic walls}

If domain walls are produced during an anisotropic phase in the early universe and are subsequently pushed outside the horizon (and freeze-out in comoving coordinates) due to inflation, they will retain the imprints of this anisotropy, which will only be erased once they re-enter the horizon and become relativistic.

\begin{figure}
\includegraphics[width=4in]{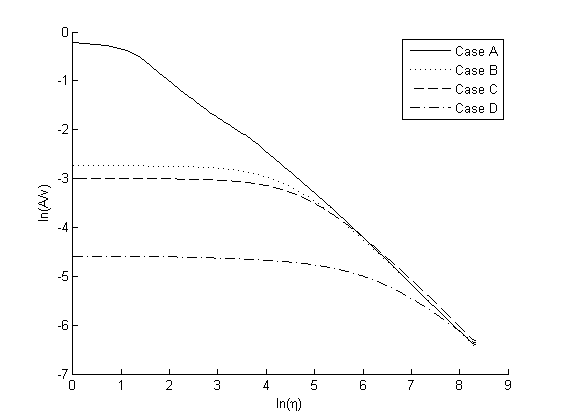}
\includegraphics[width=4in]{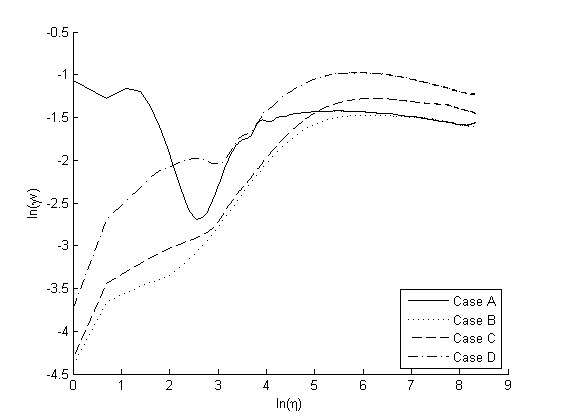}
\caption{\label{isotropization}The evolution of the network density (specifically $A/V\propto\rho$) and velocity (specifically $\gamma v$) for the four cases described in the main text. The plotted quantities are the result of averaging over ten $8192^2$ simulations, and the fits discussed in the text were done in the last ten percent of the simulation times.}
\end{figure}
\begin{figure}
\includegraphics[width=4in]{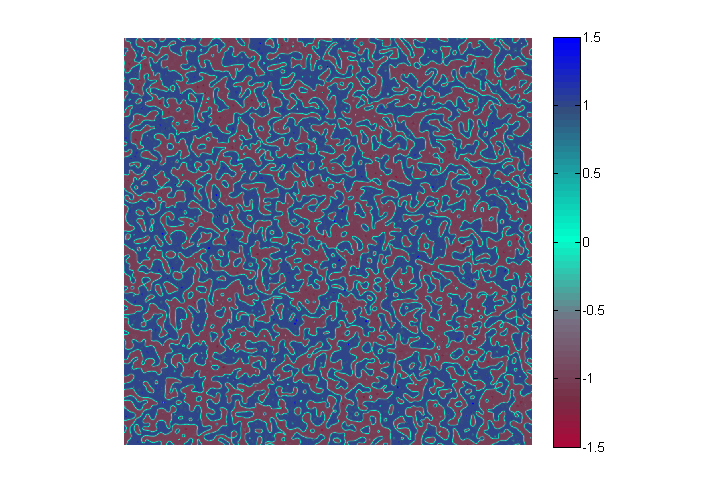}
\includegraphics[width=4in]{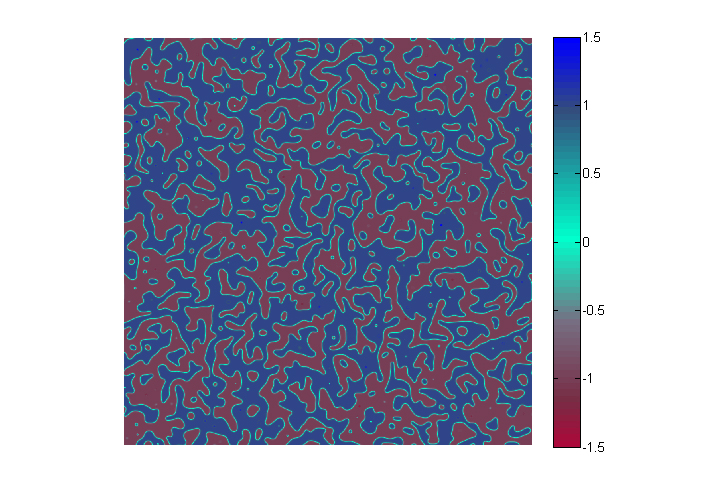}
\includegraphics[width=4in]{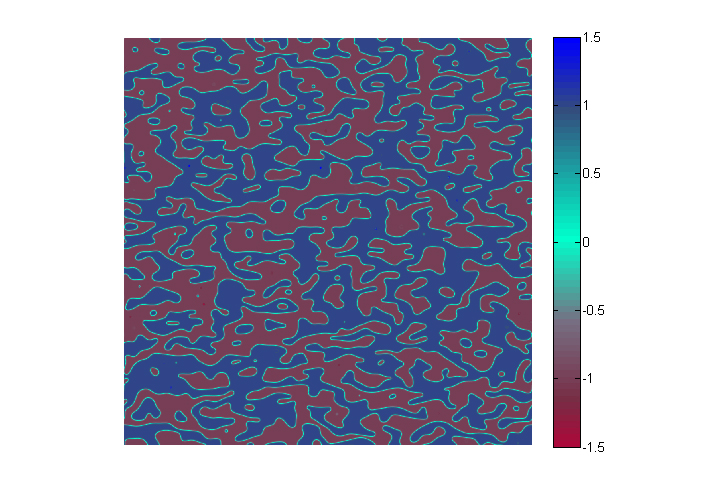}
\caption{\label{boxearly}Box snapshots of $1024^2$ simulations of cases A (top), B (middle) and C (bottom), at a conformal time $\eta=20$. The three simulations are clearly distinguishable.}
\end{figure}
\begin{figure}
\includegraphics[width=4in]{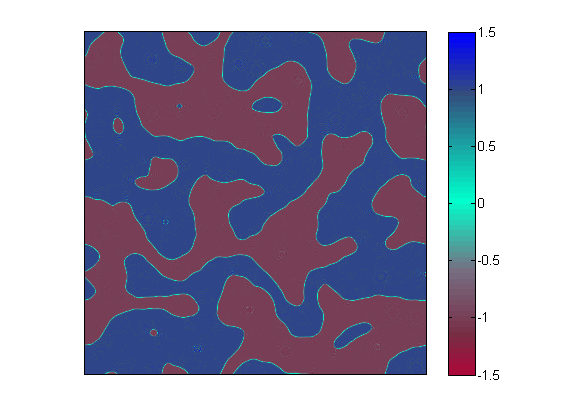}
\includegraphics[width=4in]{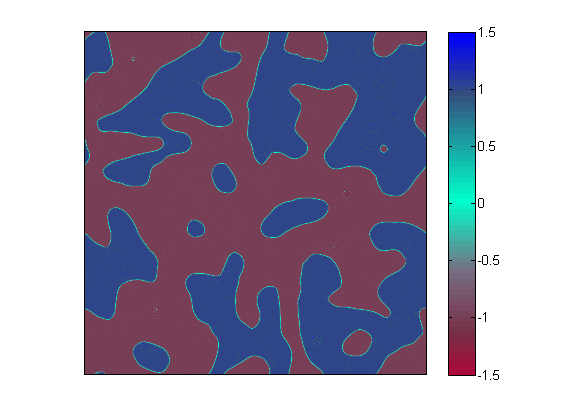}
\includegraphics[width=4in]{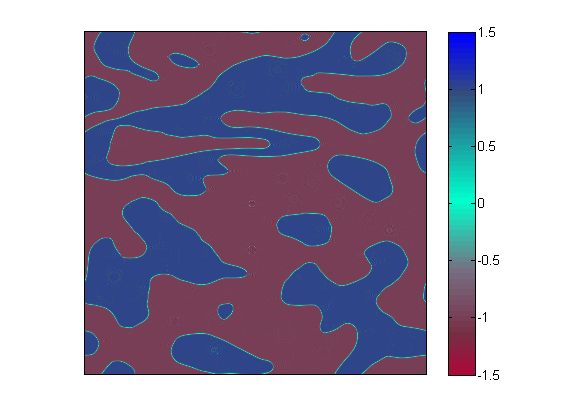}
\caption{\label{boxlate}Box snapshots of $1024^2$ simulations of cases A (top), B (middle) and C (bottom), at a conformal time $\eta=512$. Statistically the three snapshots are now quite similar.}
\end{figure}

This scenario was discussed in \cite{Fossils}, which presented some evidence for this isotropization process. However, this earlier work mostly relied on small ($1024^2$) simulations and the defect velocities were not accurately measured (nor scaling exponents for the velocity calculated). Moreover, the inferred scaling exponent, $\mu\sim-0.88$, suggested that full scaling had not been achieved---most likely due to the relatively small size of the simulations, as discussed in \cite{Leite1}. In what follows we provide more quantitative evidence for this isotropization.

Note that our interest is to study the evolution of the networks in the recent (post-inflationary) universe, so we do not need to simulate anisotropic universes, but only initially anisotropic networks evolving in an isotropic universe. In practice we need to compare three cases:
\begin{itemize}
\item Case A: Standard networks, generated with initial conditions as described in the previous section and evolving from an initial conformal time $\eta_i=1$ to a final one equal to half the box size, $\eta_f=N_{\rm box}/2$
\item Case B: Super-horizon isotropic networks, which start evolving at $\eta_i=1$  with initial conditions obtained from previous simulations at the conformal time $\eta_h=20$ and with velocities reset to zero; the choice of $\eta_h=20$ (which is twice the wall thickness) corresponds to a time where the network is reasonably well defined. This setup numerically mimics a network that is initially outside the horizon.
\item Case C: Initial conditions as in case B, but stretched in one direction by a factor $f=2$.
\item Case D: Initial conditions as in case B, but stretched in one direction by a factor $f=16$.
\end{itemize}

\begin{table*}
\begin{center}
\begin{tabular}{|c|c|c|c|c|}
\hline
Case/Box & $1024^2$ & $2048^2$ & $4096^2$ & $8192^2$ \\
\hline
 A  & $\mu=-0.95\pm0.10$ & $\mu=-0.93\pm0.10$ & $\mu=-0.94\pm0.08$ & $\mu=-0.98\pm0.08$ \\
{ } & $\nu=-0.08\pm0.24$ & $\nu=-0.08\pm0.06$ & $\nu=-0.07\pm0.05$ & $\nu=-0.06\pm0.03$ \\
\hline
 B  & $\mu=-0.85\pm0.10$ & $\mu=-0.91\pm0.09$ & $\mu=-0.95\pm0.08$ & $\mu=-0.96\pm0.07$ \\
{ } & $\nu=-0.09\pm0.14$ & $\nu=-0.03\pm0.10$ & $\nu=-0.04\pm0.03$ & $\nu=-0.05\pm0.02$ \\
\hline
 C  & $\mu=-0.85\pm0.09$ & $\mu=-0.87\pm0.13$ & $\mu=-0.97\pm0.07$ & $\mu=-0.95\pm0.08$ \\
{ } & $\nu=-0.01\pm0.17$ & $\nu=-0.07\pm0.10$ & $\nu=-0.04\pm0.03$ & $\nu=-0.06\pm0.03$ \\
\hline
 D  & $\mu=-0.39\pm0.03$ & $\mu=-0.56\pm0.06$ & $\mu=-0.71\pm0.07$ & $\mu=-0.82\pm0.09$ \\
{ } & $\nu=-0.11\pm0.15$ & $\nu=-0.09\pm0.06$ & $\nu=-0.10\pm0.03$ & $\nu=-0.12\pm0.09$ \\
\hline
\end{tabular}
\caption{Scaling exponents $\mu$ and $\nu$ (with corresponding one-sigma uncertainties), for the four cases discussed in the main text, as a function of the box size of the simulation.}
\label{tableanis}
\end{center}
\end{table*}

Fig. \ref{isotropization} shows the evolution of the densities (specifically $A/V$) and velocities (specifically, $\gamma v$, $\gamma$ being the Lorentz factor) for our three sets of simulations. In all cases we show the results of averaging over sets of ten $8192^2$ matter era simulations (in order to simplify the plots, the corresponding error bars are not displayed). Figures \ref{boxearly} and \ref{boxlate} show snapshots of the evolution of $1024^2$ boxes in each of the three cases. The snapshots in Fig. \ref{boxearly} correspond to conformal time  $\eta=20$ (cf. $\log{20}\sim3$ in Fig. \ref{isotropization}), and the differences in density are clearly visible; on the other hand, the snapshots in Fig. \ref{boxlate} correspond to the final timestep ($\eta=512$), and it is equally clear that at this stage of their evolution the three networks are quite similar.

We then used the data from the final ten percent of each set of simulations (as well as of sets of simulations with smaller box sizes) to fit for the scaling exponents $\mu$ and $\nu$ defined in the previous section; the best-fit scaling exponents (with corresponding one-sigma uncertainties) are summarized in Table \ref{tableanis}. This analysis confirms that within the statistical uncertainties the networks in cases A, B and C have reached scaling: our results are consistent with $\mu=-1$ and $\nu=0$. For the more anisotropic case D convergence to scaling is also clear, although the timescale involved is clearly longer, and therefore a larger dynamical range is needed to reach it. Indeed, itt is also noticeable that the fitted scaling exponents do change with box size, and converge as this size is increased. This explains the results of earlier studies, based on smaller simulations.

\section{\label{skewed} Biased initial conditions}

The usual choice of initial conditions assumes $\phi$ to be a random variable uniformly distributed between $-\phi_{0}$ and $+\phi_{0}$ (note that in some previous works any value between $-\phi_{0}$ and $+\phi_{0}$ is allowed, while others only allow the values $-\phi_{0}$ and $+\phi_{0}$ themselves). Thus the fraction of the simulation box that is in either minimum is initially $50\%$, and this fraction is maintained by the subsequent evolution.

\begin{figure}
\includegraphics[width=3.5in]{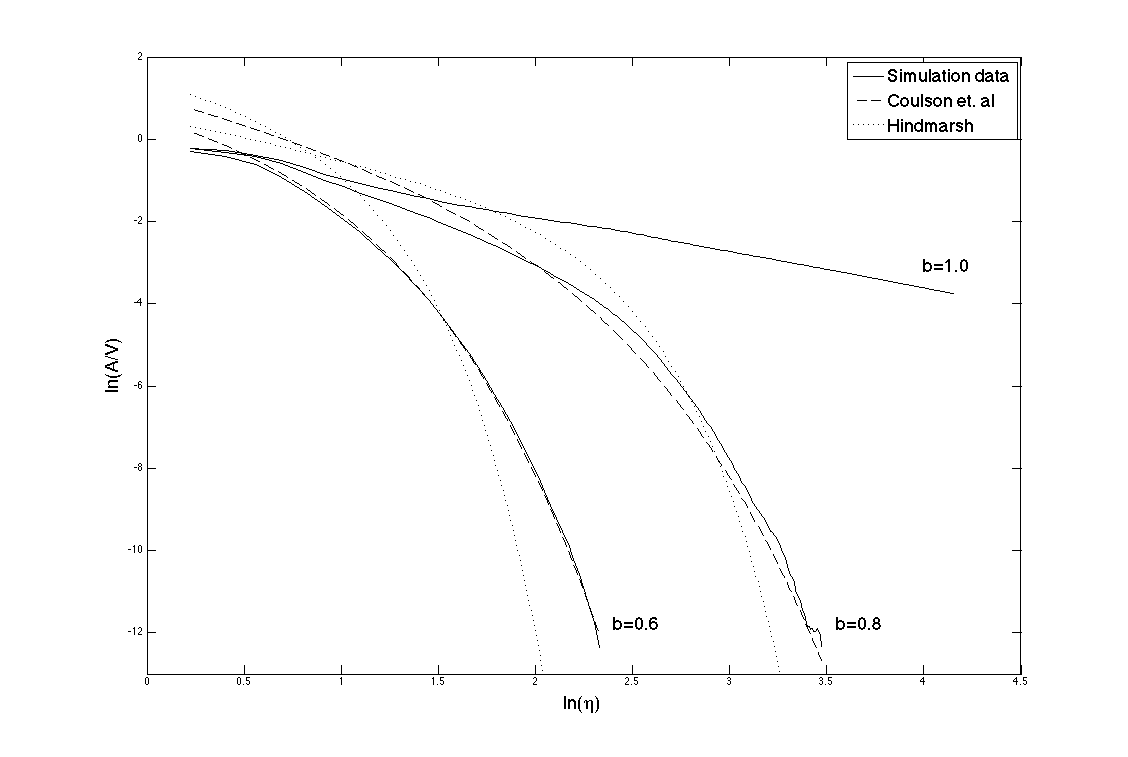}
\includegraphics[width=3.5in]{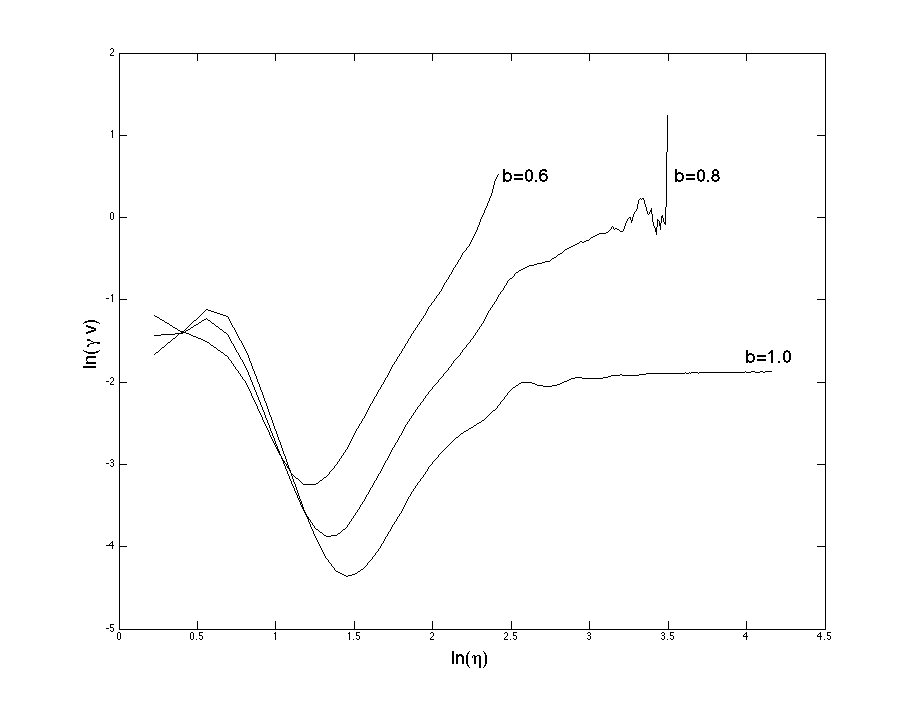}
\caption{\label{initcond}The evolution of the network density (specifically $A/V$) and velocity (specifically $\gamma v$) in the matter-dominated era for the cases $b=1.0$ (unbiased case), $b=0.8$ (weak bias) and $b=0.6$ (strongb bias) described in the main text. The plotted quantities are the result of averaging over five simulation. Also plotted are the fitting functions discussed in the main text.}
\end{figure}
\begin{figure}
\includegraphics[width=3.5in]{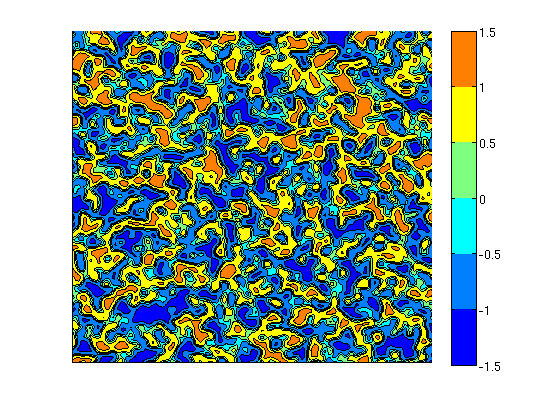}
\includegraphics[width=3.5in]{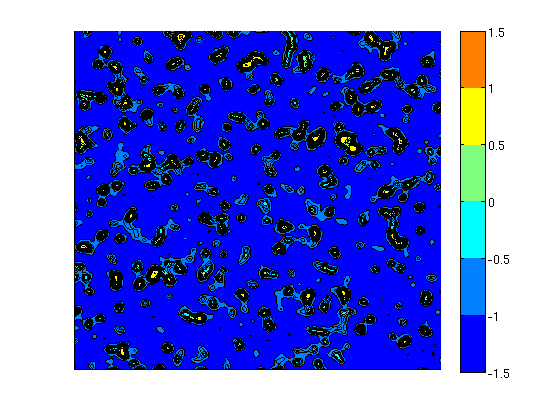}
\includegraphics[width=3.5in]{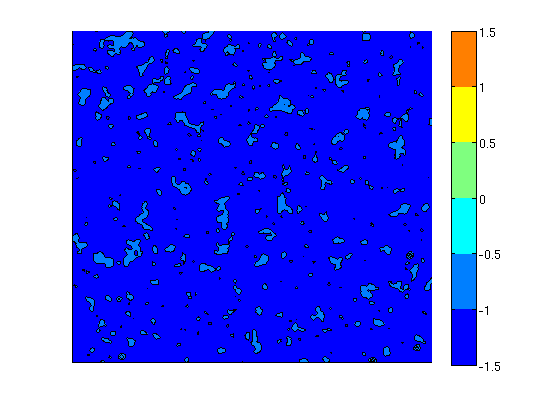}
\caption{\label{boxbias}Box snapshots of a fraction of $2048^2$ simulations of cases $b=1$ (unbiased case, top), $b=0.8$ (middle) and $b=0.6$ (bottom), at a conformal time $\ln{\eta}=2.25$ when the latter network is about to disappear.}
\end{figure}

One can, however, bias the initial conditions by changing the above fractions; a previous inflationary phase could again be responsible for this, by creating Hubble volumes with slightly different occupation fractions. In this biased case, we expect the network to eventually disappear. This situation was first considered by \cite{Coulson} (and later by \cite{Larsson}), but again this was based only on $1024^2$ boxes, and the fact that the former work obtains $\mu=-0.88\pm0.04$ for the standard (unbiased) case again suggests that the box size may be too small to study the decay process, and in particular to compare it to the analytic predictions of \cite{Hindmarsh}.

The phenomenological analysis of Coulson \textit{et al.} \cite{Coulson} suggests that for population fractions close to $50\%$ (ie, a weak bias) a good fit is provided by
\begin{equation}
\frac{A}{V}\propto\eta^{-1}\exp{\left(-\eta/\eta_c\right)}\,, \label{coulson1}
\end{equation}
while for a stronger bias
\begin{equation}
\frac{A}{V}\propto\exp{\left(-\eta/\eta_c\right)}\,, \label{coulson2}
\end{equation}
is sufficient. In both cases $\eta_c$ provides a characteristic timescale at which the decay starts. Later on, Hindmarsh \cite{Hindmarsh} provided some analytic arguments suggesting that, in the case of two spatial dimensions which we are considering in the present work, and assuming population fractions of $1/2\pm\epsilon$, one would expect to have
\begin{equation}
\frac{A}{V}\propto\eta^{-1}\exp{\left[-\kappa\epsilon^2\eta^2\right]}\,, \label{coulson3}
\end{equation}
The subsequent analysis of Larsson \textit{et al.} \cite{Larsson} claims a `reasonable' agreement with this formula, using $1024^2$ simulations and $\epsilon$ in the range $0.00--0.08$, though no quantitative measure of this agreement is provided.

We have simulated this case by generating initial conditions where the field is uniformly distributed between $-\phi_{0}$ and $+b\phi_{0}$. When $b=1$ we recover the standard (unbiased) case discussed in the previous section, while when $b=0$ the whole initial box starts on the same side of the potential and we have no domain walls. (This latter case was also confirmed numerically, as a simple test of the code.) Thus the initial population fractions in the negative and positive minima are, respectively,
\begin{equation}
f_-=\frac{1}{1+b}\,,\quad f_+=\frac{b}{1+b}\,, \label{popbias}
\end{equation}
or equivalently
\begin{equation}
\epsilon=\frac{1-b}{2(1+b)}\,. \label{ourepsil}
\end{equation}

Figure \ref{initcond} shows the results of $2048^2$ matter era simulations with values of $b=1$ (unbiased case, $\epsilon=0$), $b=0.8$ ($\epsilon=1/18$) and $b=0.6$ ($\epsilon=1/8$). In each case the plotted results correspond to an average over five simulations with different (random) initial conditions. Figure \ref{boxbias} shows snapshots of a region of size $300^2$ within one box of each case, at a time $\ln{\eta}=2.25$ when the $b=0.6$ network is about to disappear.

For the case $b=1$ we find scaling exponents consistent with the results of the previous section. For the decaying cases we find that the phenomenological formulas of Coulson \textit{et al.} provide very good fits. The fitted values for the decay timescale $\eta_c$ and the reduced chi-square of the best fit are respectively
\begin{equation}
\frac{1}{\eta_c}=0.328\pm0.002\,,\quad \chi_\nu^2=1.33 \label{fitb8}
\end{equation}
for the weak bias case $b=0.8$ and
\begin{equation}
\frac{1}{\eta_c}=1.359\pm0.002\,,\quad \chi_\nu^2=1.18 \label{fitb6}
\end{equation}
for the strong bias case $b=0.6$. On the other hand, if we fix the value of $\epsilon$ (corresponding to the value of $b$ being used) in Eq. \ref{coulson3} for the analytic formulas of Hindmarsh we find much poorer fits. Figure \ref{initcond} also depicts the best fits obtained with both the phenomenological formulas of \cite{Coulson} and the analytic approximation of \cite{Hindmarsh}. Thus it is statistically clear that the square dependence on conformal time in the exponential of the Hindmarsh fitting formula is incorrect.

\section{\label{tilted} Biased potential}

An asymmetry between the two minima of the potential can also be introduced \cite{Gelmini,Larsson,DEVAL}. In this case the volume pressure from the biasing provides an additional mechanism which will affect the dynamics of these walls. A simple tilted potential is
\begin{equation}
V(\phi)=V_0\left[\left(\frac{\phi^2}{\phi_{0}^2}-1\right)^2+\mu\frac{\phi}{\phi_{0}}\right]\,,
\end{equation}
and the asymmetry parameter (or energy difference between the two vacua) is
\begin{equation}
\delta V=2\mu V_0\,.
\end{equation}
For a network with characteristic curvature radius $R$ the surface pressure (from the tension force) is
\begin{equation}
p_T=\frac{\sigma}{R}
\end{equation}
while the volume pressure (from the energy difference between the two minima) is
\begin{equation}
p_V=\delta V\,.
\end{equation}
Depending on the relative importance of these two mechanisms, the walls may be long-lived (as in the standard case) or disappear almost immediately. At early times the surface tension tends to dominate (due to the small curvature radii), and as long as this is the case we expect a linear scaling regime as in the standard case.

On the other hand, when the domains become large enough they will decay. Thus we typically expect this to happen when
\begin{equation}
R\sim\frac{\sigma}{\delta V}\,,\label{decaybound}
\end{equation}
and assuming that $R\sim\eta$ this corresponds to
\begin{equation}
\eta\sim\frac{\phi_{0}}{\mu\sqrt{V_0}}\,.\label{decaytimes}
\end{equation}
With our numerical parameters this corresponds to
\begin{equation}
\eta\sim\frac{2.25}{\mu}\,.\label{decayformula}
\end{equation}
Once the volume pressure becomes significant the walls are expected to move with an acceleration
\begin{equation}
\frac{\delta V}{\sigma}\sim\lambda^{1/2}\mu\phi_{0}\,\label{accel}
\end{equation}
and rapidly disappear. A sufficiently fast decay may allow these networks to avoid the Zel'dovich bound \cite{ZEL}.

In Larsson \textit{et al.} \cite{Larsson}, $1024^2$ simulations with $\mu$ in the range $0.0--0.015$ were studied, and assuming the fitting function
\begin{equation}
\frac{A}{V}\propto\eta^{-1}\exp{\left[-\kappa(\mu\eta){}^n\right]}\,, \label{coulson4}
\end{equation}
they suggest that a good fit is provided by an exponent $n=2\pm1$. Note that this $n=2$ case again corresponds to the Hindmarsh fitting formula.

\begin{figure}
\includegraphics[width=3.5in]{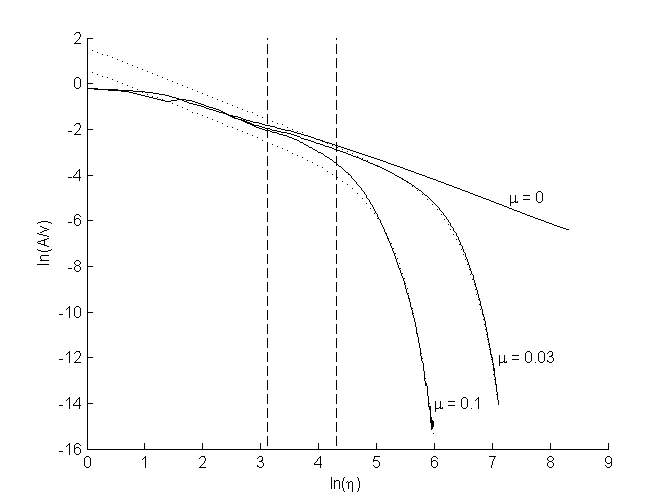}
\includegraphics[width=3.5in]{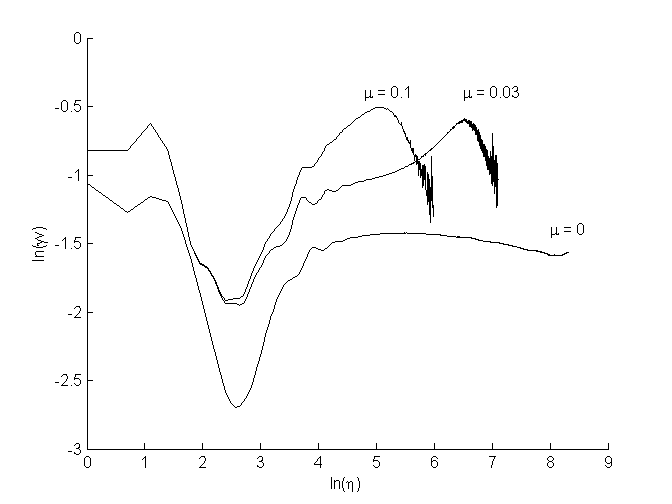}
\caption{\label{miucond}The evolution of the network density (specifically $A/V$) and velocity (specifically $\gamma v$) in the matter-dominated era for the cases $\mu=0$ (unbiased), $\mu=0.03$ and $\mu=0.1$ (dashed) described in the main text. The plotted quantities are the result of averaging over five realizations. Also plotted are the expected decay timescales discussed in the text.}
\end{figure}
\begin{figure}
\includegraphics[width=3.5in]{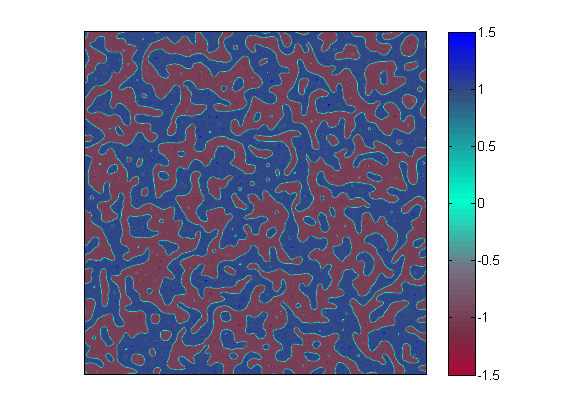}
\includegraphics[width=3.5in]{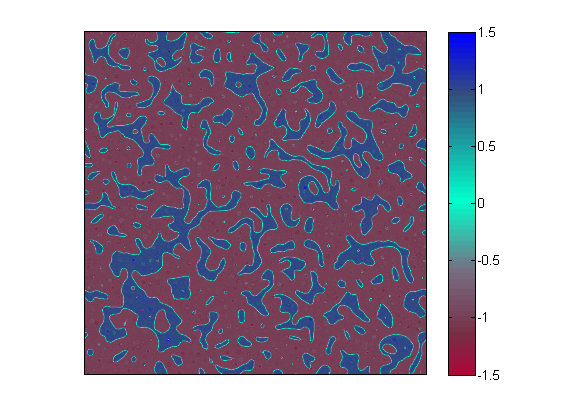}
\includegraphics[width=3.5in]{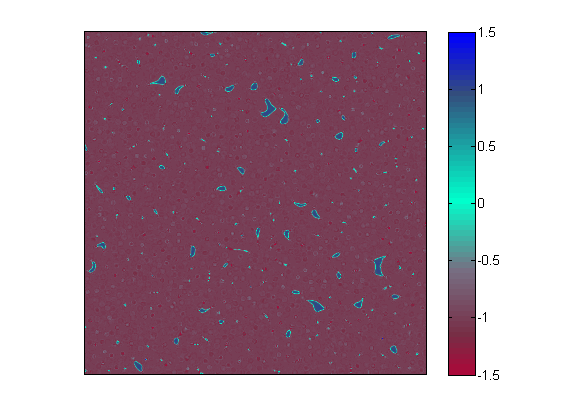}
\caption{\label{boxmiu}Box snapshots of a fraction of $2048^2$ simulations of cases $\mu=0$ (unbiased case, top), $\mu=0.03$ (middle) and $\mu=0.1$ (bottom), at a conformal time $\ln{\eta}=5$ when the latter network is about to disappear.}
\end{figure}

Figure \ref{miucond} shows the results of $2048^2$ matter era simulations with values of $\mu=0$ (unbiased case), $\mu=0.03$ and $\mu=0.1$. In each case the plotted results correspond to an average over five simulations with different (random) initial conditions. Figure \ref{boxmiu} shows snapshots of a region of size $1024^2$ within one box of each case, at a time $\eta=144$ (that is $\ln{\eta}=5$) when the more strongly biased network is about to disappear.

In this case we confirm that the choice $n=2$ suggested by Hindmarsh provides good fits; specifically for $\mu=0.03$
\begin{equation}
\kappa=(6.34\pm0.01)\times10^{-3}\,,\quad \chi_\nu^2=1.05 \label{fitmiu3}
\end{equation}
while for $\mu=0.1$
\begin{equation}
\kappa=(6.36\pm0.01)\times10^{-3}\,;\quad \chi_\nu^2=1.16 \label{fitmiu1}
\end{equation}
note that the value of $\kappa$ is the same (within the uncertainty, which is listed at the one-sigma level) in both cases. We have confirmed that other values of $n$ (say $n=1$ or $n=3$) provide much poorer fits. Figure \ref{miucond} also depicts these best-fit lines.

\section{\label{conc}Conclusions}

Several decades of analytical and numerical work on the cosmological evolution of the topological defect networks has gradually established that the so-called linear scaling solution, where the network's correlation length grows as fast as is allowed by causality and its RMS velocity is a constant, is the attractor solution for the evolution under a very broad set of scenarios. (A brief survey of some of these scenarios, and their effects on the ecolution of the defect networks, may be found in \cite{NEWTON}.) In the present work we explored a particular aspect of this issue, by studying the evolution of several types of biased domain wall networks, further quantifying whether or not the standard linear scaling solution persists and, when it doesn't, how the networks decay.

We have carried out larger numerical simulations than currently available in the literature for networks of this type, allowing for a more detailed study of the evolution of the networks. We have confirmed that anisotropic walls reach scaling, with the timescale for convergence depending on the degree of anisotropy. On the other hand for biased initial conditions or a biased potential scaling eventually breaks down and the networks decay.

We also presented, for the first time, measurements of the evolution of the averaged wall network velocities in these three scenarios. A comparison of the bottom panels of Figs. \ref{isotropization}, \ref{initcond} and \ref{miucond} provides a useful illustration of the different dynamics in each case. In the anisotropic case the networks are initially non-relativistic, but the speed gradually increases as they fall inside the horizon, and eventually becomes constant as scaling is reached. In the biased initial conditions case the defects in the decaying network become ultra-relativistic as they typically will be separating the dominant phase from collapsing bubbles of the subdominant phase. Finally in the biased potential case the velocity enhancement, although still clearly detectable in the simulations, is much smaller than in the biased initial conditions case.

More importantly, we took advantage of the larger dynamic range of our simulations to test some previously published phenomenological decay laws for networks with biased initial conditions and a biased potential. In the former case we found that the ad-hoc fitting formulas of Coulson \textit{et al.} \cite{Coulson} provide a good fit to our simulations, in contrast to the formula deduced from the analytic model of Hindmarsh \cite{Hindmarsh}. In the latter case we have improved on the results of Larsson \textit{et al.} \cite{Larsson}, confirming a scaling exponent (which was uncertain in previous work) in agreement with the Hindmarsh formula. 

Thus the above analysis leads us to conclude that the decay rate of networks with biased initial conditions differs from that of networks with a biased potential, and in particular only the latter is well described by the Hindmarsh analytic formula. The interesting follow-up question is then what is the physical reason for this difference between the two scenarios.

Although we have not studied the issue in detail, we may speculate that it could be related to Hindmarsh's choice of a Gaussian ansatz for the field probability distribution. For the biased potential case this should in principle be a good approximation, but it is not clear whether this is also the case when we have biased initial conditions. Indeed, insight from the much more extensive numerical studies of cosmic string networks may lead us to expect \cite{Fractal} it to be a good approximation in flat (Minkowski) space, but rather less so in the case of an expanding universe. An investigation of this hypothesis is beyond our present scope, and is left for future work.

\begin{acknowledgments}
This work was done in the context of the project PTDC/FIS/111725/2009 from FCT (Portugal). C.J.M. is also supported by an FCT Research Professorship, contract reference IF/00064/2012, funded by FCT/MCTES (Portugal) and POPH/FSE (EC).
\end{acknowledgments}

\bibliography{biased}
\end{document}